\documentclass{PoS}

\title{Calibrating the power of relativistic jets}

\ShortTitle{Calibrating jet power}

\author{\speaker{L. Foschini}$^{,\, a}$, M.~L. Lister$^b$, T. Hovatta$^{c,d}$, Y.~Y. Kovalev$^{e,f}$, P. Romano$^a$, S. Vercellone$^a$, A. L\"ahteenm\"aki$^{c,g}$, T.~K. Savolainen$^{g,c,f}$, M. Tornikoski$^c$, E. Angelakis$^f$, M. Berton$^{d,c}$, V. Braito$^a$, S. Ciroi$^h$, M. Kadler$^i$, P.~R. Burd$^i$\\
        \llap{$^a$} Istituto Nazionale di Astrofisica (INAF), Osservatorio Astronomico di Brera, Merate (Italy), E-mail: \email{luigi.foschini@inaf.it}.\\
        \llap{$^b$} Purdue University, Department of Physics and Astronomy, West Lafayette (USA)\\
        \llap{$^c$} Aalto University, Mets\"ahovi Radio Observatory, Kylm\"al\"a (Finland)\\
        \llap{$^d$} Finnish Centre for Astronomy with ESO (FINCA), Turku (Finland)\\
        \llap{$^e$} Astro Space Center Lebedev Physical Institute, Moscow (Russia)\\
        \llap{$^f$} Max-Planck-Institut f\"ur Radioastronomie, Bonn (Germany)\\
        \llap{$^g$} Aalto University, Department of Electronics and Nanoengineering, Espoo (Finland)\\
        \llap{$^h$} University of Padova, Department of Physics and Astronomy, Padova (Italy)\\
        \llap{$^i$} Universit\"at W\"urzburg, Institut f\"ur Theoretische Physik und Astrophysik, W\"urzburg (Germany)
        }

\abstract{There are several methods to calculate the radiative and kinetic power of relativistic jets, but their results can differ by one or two orders of magnitude. Therefore, it is necessary to perform a calibration of the jet power, to understand the reasons for these differences (whether wrong hypotheses or intrinsic source variability), and if it is possible to converge to a reliable measurement of this physical quantity. We present preliminary results of a project aimed at calibrating the power of relativistic jets in active galactic nuclei (AGN) and X-ray binaries (XRB). We started by selecting all the AGN associations with known redshift in the Fourth \emph{Fermi} LAT Gamma-Ray Catalog (4FGL). We then calculated the radiative and/or kinetic powers from available data or we extracted this information from literature. We compare the values obtained for overlapping samples and highlight preliminary conclusions.}

\FullConference{High Energy Phenomena in Relativistic Outflows VII - HEPRO VII\\
		9-12 July 2019\\
		Facultat de F\'isica, Universitat de Barcelona, Spain}

\begin{document}

\section{Introduction}
The measurement of the power, either kinetic or radiative, emitted by relativistic jets is one of the fundamental tasks required to understand the nature of jetted sources. A direct method is not available, but there are several ways to calculate it starting from observed quantities. However, results from different methods can differ by one or more orders of magnitude. Therefore, it is necessary to understand if these differences are due, for example, to wrong hypotheses adopted in the calculations, or to the intrinsic variability of the sources. This is the aim of the present project and we present here some preliminary results. Only some attempts of this type of comparison have been published to date \cite{FOSCHINI,SHABALA12,SHABALA13,GODFREY13,GODFREY16,PJANKA}. Here we would like to perform a more systematic and detailed study on a larger sample and on more recent observations. This essay refers to early comparisons on samples of AGN. X-ray binaries (XRB) will be studied in further works.

All the powers and luminosities were calculated with reference to a $\Lambda$CDM cosmology with $H_{0}=70$~km~s$^{-1}$~Mpc$^{-1}$ and $\Omega_{\rm m}=0.3$. A few authors adopted $H_{0}=71$~km~s$^{-1}$~Mpc$^{-1}$: in these cases, we recalculated the proper values. 

\section{Methods selection}
There are many ways to calculate the jet power. After a careful examination of the available literature, we decided to focus on the following four methods:

\begin{enumerate}
\item The simplest way to estimate the radiative power is to use {\bf the luminosity measured in high-energy $\gamma$ rays}, the Doppler and the bulk Lorentz factors: $P_{\rm rad}\sim \Gamma^2(L_{\rm \gamma}/\delta^4)$. This is a first-order estimation, as the radiative power depends on the radiative process (external Compton \emph{versus} Synchrotron Self-Compton), the type of jet (continuous \emph{versus} blobs), viewing angle, and other parameters (see e.g., \cite{BLANDFORD,SIKORA,GHISELLINI}). The \emph{Fermi} Large Area Telescope (LAT) Collaboration offers periodic catalogs of all the sources detected by LAT in the $(0.1-300)$~GeV energy range. Recently, the fourth version of this catalog has been released \cite{FERMI}. Doppler and bulk Lorentz factors can be measured by means of radio observations. The Doppler factor $\delta$ can be measured by means of the flux decay of the knots as observed with the VLBA (Very Long Baseline Array) at 43~GHz \cite{JORSTAD}, by studying the variability brightness temperature via flux density variations \cite{LIODAKIS}, or by solving a system of equations where the basic observed quantities are the jet flux and the radio core shift \cite{FINKE}. Another important quantity that can be measured directly is the apparent speed of a knot $\beta_{\rm app}$, by means of high-angular resolution radio observations \cite{LISTER16,JORSTAD,PUSHKAREV,LISTER19}. Once $\delta$ and $\beta_{\rm app}$ are known, it is possible to calculate the bulk Lorentz factor according to the formula $\Gamma = (\beta_{\rm app}^2+\delta^2+1)/(2\delta)$. The distributions of $\Gamma$ for the sources in common between the different samples are displayed in Fig.~\ref{fig1}, which should be compared with the expected distributions for flux limited jet samples calculated by \cite{LISTER19} (see Fig.~11 of that work). It is worth noting that often it is not possible to have simultaneous measurements of $\delta$ and $\beta_{\rm app}$, which implies the appearance of anomalous values of $\Gamma$ (cf. Fig.~\ref{fig1}). The only work making use of simultaneous measurements is that of Jorstad et al. \cite{JORSTAD}, which uses frequent cadence 43~GHz VLBA observations to track the kinematics and brightness temperature of rapidly fading jet features near the jet core. The $15$~GHz \emph{Monitoring Of Jets in Active galactic nuclei with VLBA Experiments} (MOJAVE) program \cite{LISTER19} observes a much larger sample of AGN jets, but with insufficient angular resolution and monitoring cadence to enable such an analysis.

\begin{figure}[t]
\centering
     \includegraphics[width=.45\textwidth]{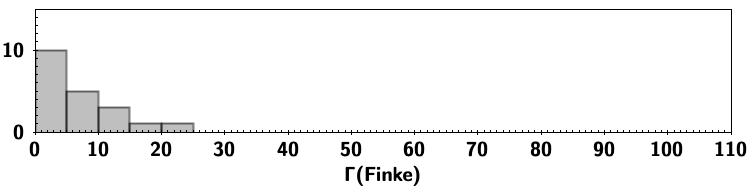}
     \includegraphics[width=.45\textwidth]{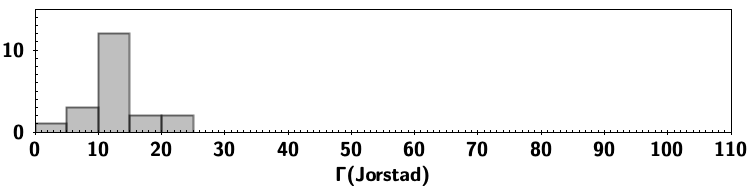}\\
     \includegraphics[width=.45\textwidth]{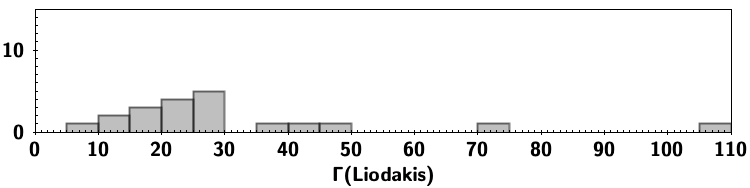}
     \includegraphics[width=.45\textwidth]{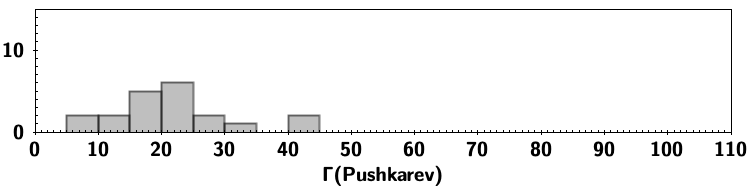}
     \caption{Distribution of $\Gamma$ for the sources in common between the different samples of \cite{FINKE,JORSTAD,LIODAKIS,PUSHKAREV}.}
     \label{fig1}
\end{figure}

\item Another widely adopted method is {\bf the modeling of the spectral energy distribution (SED)} by using simple one-zone leptonic models \cite{GHISELLINI,PALIYA17,PALIYA19}. Hadronic models are generally disfavored, and, therefore, we do not consider them for comparison. SED modeling can disentangle different contributions to the jet kinetic power, such as the magnetic field, electrons, and protons. The kinetic power carried by protons is generally calculated by assuming one baryon per lepton, but this hypothesis was barely tested and there are reasons to believe it may not be appropriate \cite{PJANKA}. The radiative power is calculated according to the formula $P_{\rm rad}=2fL_{\rm bol}/\Gamma^2$, where $f$ is a numerical factor depending on the radiative process ($16/5$ for synchrotron and synchrotron self-Compton, $4/3$ for external Compton), and the adopted formula is valid at a critical viewing angle such that $\delta \sim \Gamma$. The factor $2$ takes into account that we are observing only one side of a bipolar jet (again, if the viewing angle is so small that the de-boosted counter-jet flux can be neglected). 

\item A third method is based on {\bf the cavities excavated by the jet in the intergalactic medium} \cite{BIRZAN04,BIRZAN08,CAVAGNOLO}. In this case, it is possible to calculate the kinetic power by measuring the extended radio emission at low frequencies, in the MHz frequency range, according to the formula: $\log P_{\rm kin} = 0.64(\log L_{\rm 200-400 MHz}-40) + 43.54$ \cite{MEYER,NOKHRINA}. As the cavity formation is a process requiring billion of years, the resulting kinetic power is likely to be lower than the values obtained with other methods based on data with shorter time scales.

\item The fourth method is based on {\bf the relationship between the jet power and the luminosity of the radio core} according to the Blandford \& K\"onigl model \cite{BLANDFORD}. We used the radio fluxes measured by the MOJAVE program \cite{LISTER16,LISTER19}), while the powers are calculated according to the formulas \cite{FOSCHINI14}: $\log P_{\rm rad}=12+0.75\log L_{\rm 15GHz}$ and $\log P_{\rm kin}=6+0.90\log L_{\rm 15GHz}$. Given the rapid flux variability exhibited by radio cores due to frequent emergence of new jet features, this method is strongly influenced by source variability.
\end{enumerate}

\section{Sample selection and comparison}
\begin{figure}[t]
\centering
     \includegraphics[width=.8\textwidth]{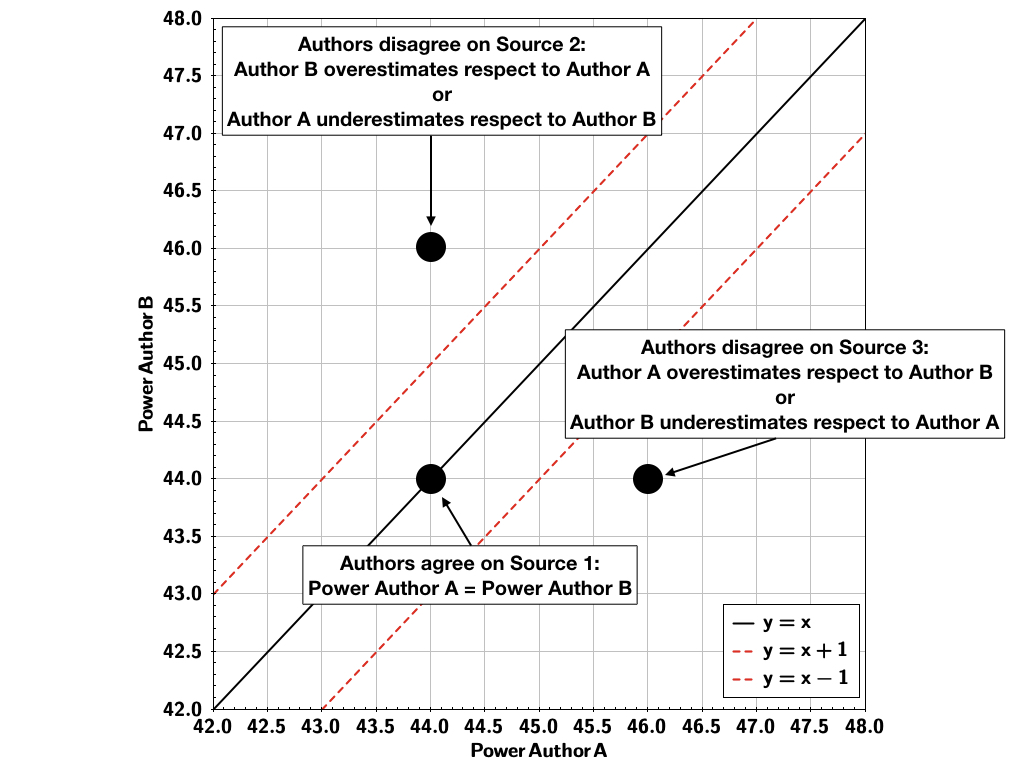}
     \caption{Explanation of the comparison plots. The figures we show are not the classical figures showing a correlation between two different quantities. In the present case, we display the same quantity as measured by two different authors. Therefore, what is important is the position of the points in the plot. Points on the line $y=x$ ($1:1$ ratio) indicate that the two measurements are consistent. Deviations from the $1:1$ ratio line indicate problems of consistency. The scale is logarithmic, so that -- for example -- the lines $y=x\pm1$ indicate one order of magnitude difference with respect to the $1:1$ ratio.}
     \label{fig2}
\end{figure}

Each of the works we selected made use of its own sample of sources. These samples have different selection criteria and are not always fully overlapping. The smallest sample is Jorstad et al.'s \cite{JORSTAD}, comprising 36 sources (21 quasars, 12 BL Lac objects, and 3 radio galaxies). The largest one is that of Paliya et al. \cite{PALIYA17,PALIYA19}, made up of 531 sources, including 16 jetted narrow-line Seyfert 1 galaxies. When comparing two methods, we always select the subsample of sources in common between the two works. It is worth stressing that we are not searching for correlations. Therefore, we are simply showing the same quantity (either the radiative or the kinetic power of the jet) for the same sources as estimated by different authors\footnote{In the present work, we present only comparisons of the same method by different authors, except in one case (Fig.~\ref{fig4}, right panel). Comparisons of different methods will be addressed in other works.}. We illustrate our method graphically in Fig.~\ref{fig2}.  

To make the comparison more quantitative, we also fit the points to a line $y= C + \alpha\cdot x$. The slope $\alpha$ indicates the relative accuracy between the two methods, while $C$ is a constant. The dispersion of the points, $\sigma$, is the relative precision. The ideal case would be $\alpha \sim 1$ and $\sigma \rightarrow 0$ (at least as small as possible, as some intrinsic variability due to data from different epochs remains).

\subsection{Comparisons of SED modeling}
The SED modeling method was adopted by Ghisellini et al. (214 sources, \cite{GHISELLINI}) and Paliya et al. (531 sources, \cite{PALIYA17,PALIYA19}). The two samples have 99 sources in common. The datasets also differ with reference to high-energy $\gamma$-rays, which are the dominant quantity in determining the jet power, Ghisellini et al. made use of the first and second \emph{Fermi} LAT catalogs, covering a time period of the first two years operations (2008-2010). Paliya et al. made their own analysis of LAT data, covering 8 years for \cite{PALIYA17} and 9 years for \cite{PALIYA19}. Therefore, a certain difference between the results is expected, randomly distributed, due to the intrinsic variability of the sources. 

\begin{figure}[t]
\centering
     \includegraphics[width=.45\textwidth]{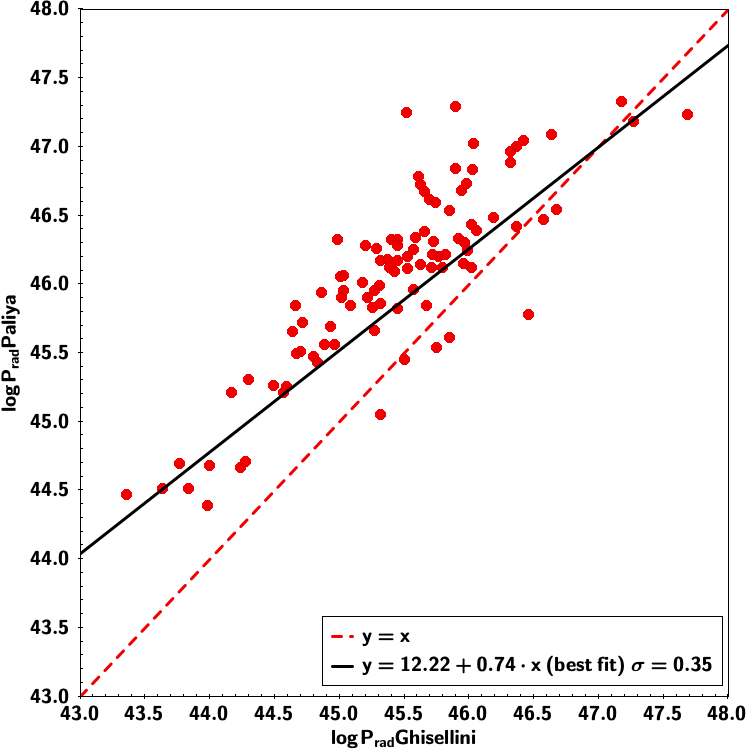}
     \includegraphics[width=.45\textwidth]{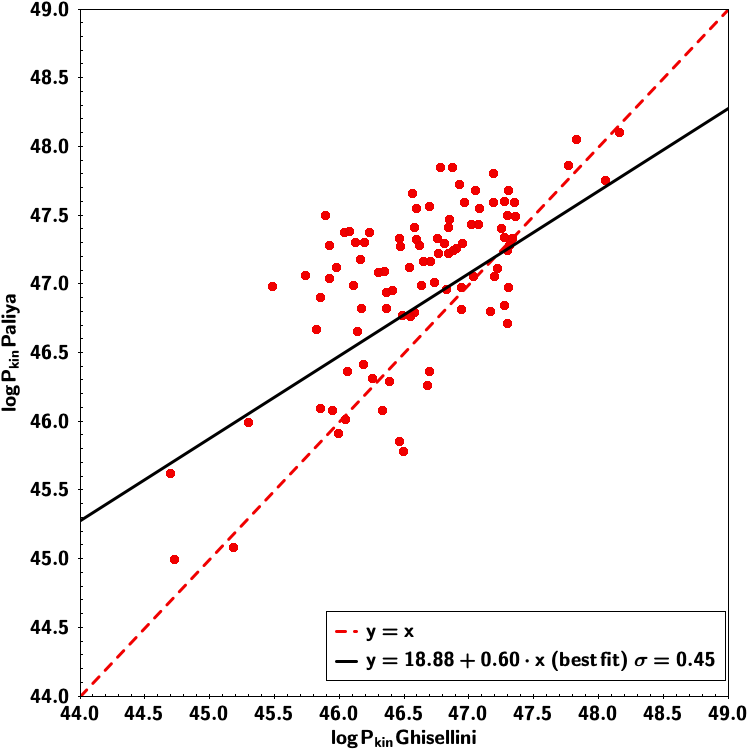}
     \caption{Comparison of jet powers (\emph{left panel}, radiative; \emph{right panel}, kinetic) from SED modeling by Ghisellini et al \cite{GHISELLINI} and Paliya et al. \cite{PALIYA17,PALIYA19}. The dashed line is the $1:1$ ratio. The continuous line is the best fit, whose $\alpha$ and $\sigma$ are indicated in the legends.} 
     \label{fig3}
\end{figure}

However, the comparison shown in Fig.~\ref{fig3} displays systematic differences between the values obtained by the two teams, despite of the adoption of the same model. The relative accuracy is $\alpha=0.74$ for the radiative power, and $\alpha=0.60$ for the kinetic one, while the dispersion is relatively contained ($\sigma=0.35$, and $\sigma=0.45$, respectively). As the authors declare to use the same model, it is not immediately clear the reason of this discrepancy. The intrinsic variability can account for $\sigma$, but not for $\alpha$. Further studies are needed.

\subsection{Comparison of kinetic powers from cavities in the intergalactic medium}
\begin{figure}[t]
\centering
     \includegraphics[width=.45\textwidth]{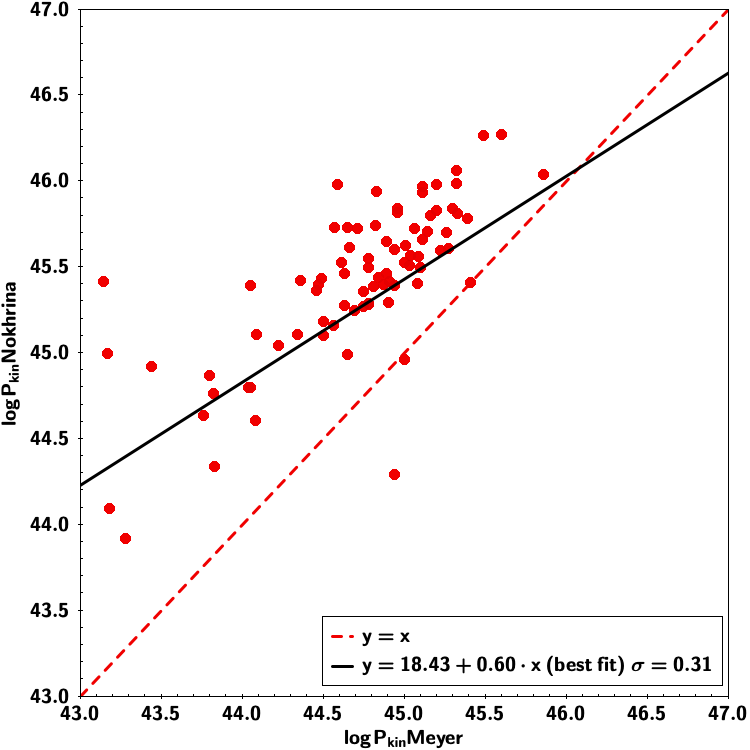}
     \includegraphics[width=.45\textwidth]{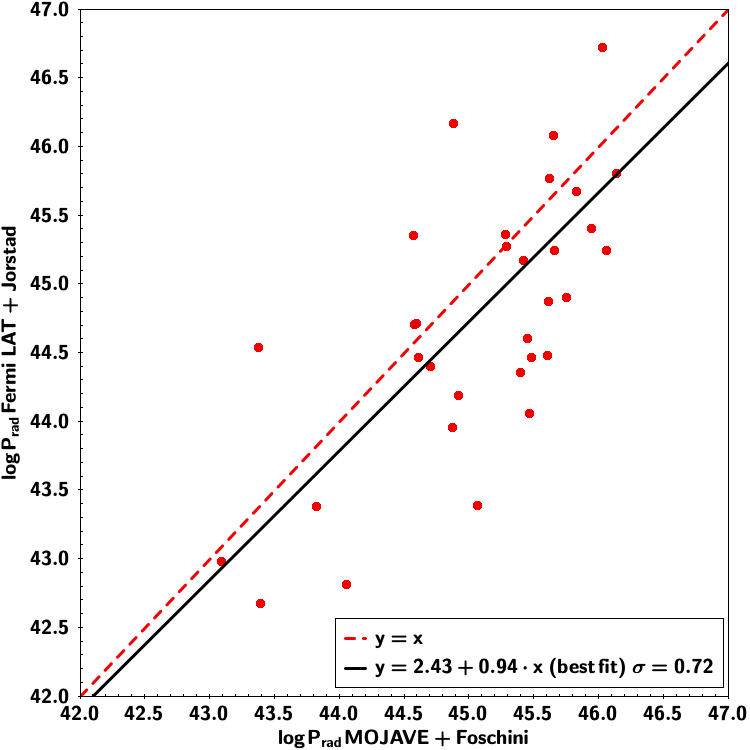}
     \caption{(\emph{left panel}) Comparison of kinetic power from extended radio emission by Meyer et al. \cite{MEYER} and Nokhrina et al. \cite{NOKHRINA}. (\emph{right panel}) Comparison of radiative power from 15~GHz radio core emission from the MOJAVE program \cite{LISTER19} plus the Foschini formula \cite{FOSCHINI14} and $\gamma-$ray luminosity from \emph{Fermi} LAT observations \cite{FERMI} plus Doppler and bulk Lorentz factors from VLBA 43~GHz observations by Jorstad et al. \cite{JORSTAD}. The dashed line is the $1:1$ ratio. The continuous line is the best fit, whose $\alpha$ and $\sigma$ are indicated in the legends.} 
     \label{fig4}
\end{figure}

The second method under examination is the kinetic power estimated from the extended radio emission \cite{BIRZAN04,BIRZAN08,CAVAGNOLO}. The sample of Meyer et al. is made of 216 sources \cite{MEYER}, while that of Nokhrina et al. has 97 entries \cite{NOKHRINA}. The resulting overlapping subsample consists of 81 sources. Again, despite the adoption of the same method, the comparison shows systematic differences (Fig.~\ref{fig4}, left panel). In this case, the reason could be the different way to estimate the extended radio emission: Meyer et al. calculated the 300~MHz ($\lambda=1$~m) luminosity by extrapolating radio data at higher frequencies, while Nokhrina et al. directly searched in different catalogs for radio measurements at 326~MHz ($\lambda=92$~cm).

\subsection{Comparison of radiative powers from radio core emission and $\gamma$-ray luminosity}
The last comparison is the only one between two different methods. On the one side, we selected the $\gamma-$ray luminosity from the Fourth \emph{Fermi} LAT catalog \cite{FERMI} coupled with the Doppler and bulk Lorentz factors as measured from VLBA radio observations at 43~GHz \cite{JORSTAD}, because these latter values are the only available ones measured simultaneously. On the other side, we have the relationships between the jet power and the radio core emission \cite{FOSCHINI}, and the VLBA 15~GHz radio observations of the MOJAVE program \cite{LISTER16,LISTER19}. The former sample is made of 36 sources, while the latter has 492 entries. The subsample we can examine is made of 33 sources. This time we find an acceptable relative accuracy ($\alpha=0.94$), but with a significant dispersion ($\sigma=0.72$). This means that the two methods are equivalent, and the differences are likely due to the intrinsic variability of the sources, given also the different frequencies adopted for the observations. 

\section{Final remarks}
What we have shown is just a selection of the preliminary results of our study. It is important to note that these comparisons confirm that the adoption of the same method is not a guarantee of consistent results. It is necessary also to take care of the adopted data set (direct measurements, extrapolations, different epochs, etc.). Differences of up to 2-3 orders of magnitude have been observed, and the intrinsic variability is not enough to explain these discrepancies. In the case of the extended radio emission, it was shown that the extrapolation from high to low frequencies observations is misleading. Significant systematic differences in the SED modeling results have also been found and require further investigation.

\end{document}